\documentclass[runningheads]{svmult}

\usepackage{makeidx}   
\usepackage{graphicx}  
\usepackage{subeqnar}  
\usepackage{multicol}  
\usepackage{physprbb}  
\makeindex             

\begin{document}
\title*{The {\it W}--function applied to the age of \protect\newline Globular Clusters}
\toctitle{The {\it W}--function applied to the age of 
\protect\newline Globular Clusters}
%
%
\titlerunning{The {\it W}--function applied}
%
\author{Miriam Rengel\inst{1,2}
\and Gustavo Bruzual\inst{2}}
\authorrunning{Miriam Rengel and Gustavo Bruzual}
%
%
\institute{TLS Tautenburg, 07778 Tautenburg, Germany
\and Centro de Investigaciones de Astronom\'{\i}a (CIDA), 
     M\'{e}rida 5101-A, Venezuela
     }

\maketitle              

\begin{abstract}
We present a statistical approach for estimating the age of Globular Clusters\index{estimation of the age of Globular Clusters} by measurement of the
likelihood between the observed cluster sequences in the Colour--Magnitude Diagram and
the synthetic cluster sequences computed from stellar evolutionary models. In the conventional 
isochrone fitting procedure, the age is estimated in a subjective way. Here, we calculate the degree of likelihood by applying
a modern statistical estimator,  
the Saha estimator {\it W}, and the interval of confidence from $\chi^{2}$ statistics. We apply this
approach to sets of three different evolutionary models. Each of these sets consists of different
chemical abundances, ages, input physics, etc. Based on our approach, we estimate
the age of NGC 6397\index{NGC 6397}, M92\index{M92} and M3\index{M3}. With a confidence level of 99\%, we find that the best estimate of the
age is 14.0 Gyr within the range of 13.8 to 14.4 Gyr for NGC 6397, 14.75 Gyr within the range of 14.50
to 15.40 Gyr for M92, and 16.0 Gyr within the range of 15.9 to 16.3 Gyr for M3.  
\end{abstract}

\section{Introduction}
Since the Globular Clusters (GCs) contain the oldest objects for which age estimates are available, they
have been recognized as being of key importance for deriving the age of the Universe. In some conventional
procedures for determining the age of GCs, like the isochrone fitting, the stellar evolutionary model has been selected in a subjective way. There, the observed position of the stars in the Colour--Magnitude Diagram (CMD) 
is compared with the position calculated from theoretical models. The model which minimizes the discrepancies between the observed and
calculated sequences gives the estimated age of the cluster. Motivated by these ideas, and due to the refinement of theoretical stellar
models and to a growing number of observations, we present a statistical 
approach~\cite{r2001} that allows to estimate the age in a more objective way by using the Saha statistics {\it W}~\cite{s1998} and $\chi^{2}$~\cite{p1986}.
As an example, we subsequently derive the age of NGC 6397, M92 and M3.

\section{Observational data and stellar evolutionary models}

\begin{figure}[t]
\begin{center}
\includegraphics[width=.53\textwidth]{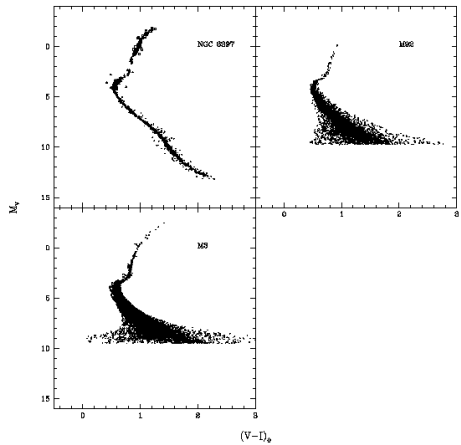}
\end{center}
\caption[]{CMDs of NGC 6397, M92 and M3, after the procedure of removing stars
and considering the adopted parameters given in~\cite{r2000} }
\label{fig1}
\end{figure}

The observational data consist of V and I photometry of three GCs of the Galaxy: NGC 6397~\cite{d1999}~\cite{k1999}, M92 and M3~\cite{r1999}.
We removed stars with excessive photometric errors (3$\sigma$ in the colour) and we selected only stars out to a radius of r$<$140' for
M92 and 170' for M3. After the procedure of rejecting stars, the sample of M92 comprises 4846 stars and the sample of M3 10333 stars
(Fig.~\ref{fig1}).

We selected the models computed by~\cite{b1994} and~\cite{g1996} (Padova isochrones), tracks developed
by~\cite{d1996} (Yale isochrones) and models by~\cite{c1998}~\cite{c1999} (Pisa isochrones). We chose
them as they were published in the theoretical plane (HRD). Every model is computed for different input data (metallicities, ages, opacities,
stellar atmospheres, etc.), resulting in differences between the isochrones. On the other hand, every model uses different forms
of conversion between the theoretical and the observational plane. Because this lack of uniformity may influence the estimation of the age, we limited this additional uncertainty
by using the same transformation to each set of isochrones: tabulations of bolometric corrections and colours determinated from the SED
extracted from the spectral library of Kurucz~\cite{b2000}.

\section{The method and the age of NGC 6397, M92 and M3}
Let S and M be the number of observed stars and synthetic stars generated from a particular stellar model in CMDs of a GC. 
If both CMDs are binned in a grid of B bins, the probability of having gathered the data under a particular model
is the likelihood function, which is given in~\ref{eq1}. By $m_{i}$ we denote the number of synthetic stars placed on the $i^{th}$ bin and
by $s_{i}$ the number of observed stars inside of the $i^{th}$ bin.
%

\begin{figure}[t]
\begin{center}
\includegraphics[width=.55\textwidth]{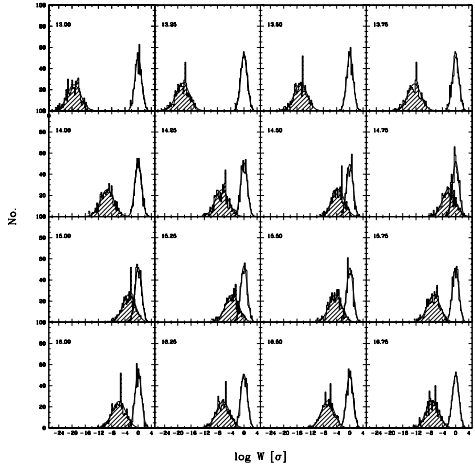}
\end{center}
\caption[]{Example of the distribution of values of log {\it W} for different
ages of the Padova isochrones for a sample of 1187 stars in NGC 6397
(see text for details)}
\label{fig2}
\end{figure}

Since there is not a unique configuration of a synthetic CMD of a GC, we generate 1000 different synthetic configurations of the CMD originating
from every isochrone of age $t_{i}$. Subsequently we draw the observational CMD and the synthetic CMDs within a grid of B cells. We then
count the number of stars $m_{i}$ in each cell for every CMD. In the next step we first calculate log {\it W} between two synthetic
CMDs (generated from the same isochrone of age $t_{i}$). Secondly, we estimate log {\it W} for a new couple of different synthetic samples
and continue determining the value of log {\it W} between different samples until 500 couples are completed (model--model comparison).
The values computed of log {\it W} give a distribution of log {\it W}, which is represented as the non--shaded region in the histogram (Fig.~\ref{fig2}).

We calculate the model--model distribution to compare it with the distribution of frequencies of log {\it W} determinated from comparing observational and synthetic samples (data--model comparison).
The data--model distributions are constructed
by computing log {\it W} for couples between 500 synthetic samples (from the same isochrone of age $t_{i}$) and the observational
data. In this way 500 values of log {\it W} are computed. The distribution of log {\it W} is shown as the shaded area in the histogram of 
Fig.~\ref{fig2}. 

We estimate the interval of confidence by using the $\chi^{2}$ statistical test, as suggested by~\cite{p1986}, between the
observational and synthetic data and for every evolutionary model. We chose the lowest value of $\chi^{2}$, $\chi^{2}_{min}$. The quantity
increases when we move away from the best fit. The values of the confidence levels for a parameter and its respective
probabilities (in percent) for data distributed in form of a Gaussian curve were taken from the statistical tables of~\cite{p1986}. 
The estimate of the age is given by the age of the isochrone that corresponds to the minimal
distance obtained between the Gaussian median fits to the model--model and data--model distributions (14.75 Gyr in Fig.~\ref{fig2}).
The full details of the approach are presented in~\cite{r2000}.

\begin{equation}
\mbox{prob} (\mbox{data} \mid \mbox{model}) = \frac{S!(M+B-1)!}{(M+S+B-1)!} \cdot W \label{eq1}
\end{equation}
where {\it W} is given by
\begin{equation}
W = \prod^{B}_{i=1} \frac{(m_i+s_i)!}{m_i!s_i!} = \prod^{B}_{i=1} \,{m_i+s_i\choose s_i}\ \label{eq2}
\end{equation}

\begin{table}[]
\caption{Summary of the results of the best estimate of the age of the
samples
considered in this work, for every evolutionary model}
\begin{center}
{\scriptsize
\begin{tabular}{cccccc}

Evolutionary & Z of the & $\chi^{2}_{min}$~~~~ & $\chi^{2}_{red}$ & Best estimate&Interval of 99\%\\
Model  & Isochrone  &  &  & of t [Gyr] &of confidence\\
\noalign{\smallskip}
\hline
\noalign{\smallskip}
NGC 6397        &1187 stars      &B=14400&      &         &
       \\[0mm]
Padua           &0.0004          &1352.81& 2.82 &14.50    & [14.3 - 15.1]     \\[0mm]
Yale            &0.0004          &1089.04& 2.34 &14.00    & [13.8 - 14.3]     \\[0mm]
\hline
NGC 6397        &373 stars      &B=8400  &      &         &
       \\[0mm]
Padua           &0.0004          &232.84 & 1.41 &14.25    & [13.7 - 15.6]     \\[0mm]
Yale            &0.0004          &206.80 & 1.25 &14.00    & [13.3 - 15.1]     \\[0mm]
Pisa            &0.0002          &249,95 & 1.41 &13.00    & [11.9 - 14.2]     \\[0mm]
\hline
M92             &4846 stars     &B=12000 &      &         &
      \\[0mm]
Padua           &0.0001         &2221.63 & 1.31 &14.75    & [14.5 - 15.4]     \\[0mm]
Yale            &0.0002         &2455.26 & 1.42 &15.00    & [14.6 - 15.7]     \\[0mm]
\hline
M92             &1482 stars     &B=12000  &      &         &
       \\[0mm]
Padua           &0.0001         &678.79  & 1.73 &15.00    & [14.8 - 15.5]     \\[0mm]
Yale            &0.0002         &769.67  & 1.91 &15.00    & [14.7 - 15.5]     \\[0mm]
Pisa            &0.0002         &832.86  & 2.06 &12.00    & [11.8 - 12.1]     \\[0mm]
\hline
M3              &10333 stars    &B=14400  &      &         &
       \\[0mm]
Padua           &0.0004         &3558.38 & 1.59 &15.75    & [15.7 - 15.9]     \\[0mm]
Yale            &0.0004         &3037.01 & 1.37 &16.00    & [15.9 - 16.3]     \\[0mm]
\hline
M3              &4929 stars     &B=14400  &      &         &
       \\[0mm]
Padua           &0.0004         &1038.53 & 2.14 &16.25    &  [16.2 - 16.6]     \\[0mm]
Yale            &0.0004         &757.20  & 1.68 &16.00    &  [15.9 - 16.3]     \\[0mm]
Pisa            &0.0002         &1120.65 & 2.45 &14.00    &  [13.9 - 14.2]     \\[0mm]
\noalign{\smallskip}
\hline
\noalign{\smallskip}
\end{tabular}}
\end{center}
\end{table}

%

\end{document}